\documentclass[prd,showpacs,twocolumn,
amsmath,amssymb,superscriptaddress,floatfix,nofootinbib]{revtex4-1}
\usepackage[utf8]{inputenc}
\usepackage[sort&compress]{natbib}
\usepackage[normalem]{ulem}
\usepackage{lipsum} 
\usepackage{bm}
\usepackage{times}
\usepackage{amssymb,amsbsy,amsmath,amsfonts}
\usepackage{graphicx}
\usepackage{float}
\usepackage{color}
\usepackage{morefloats}
\usepackage{rotating}
\usepackage{srcltx}
\usepackage{slashed}
\usepackage{subfigure}
\usepackage{multirow}
\usepackage{verbatim}
\usepackage{hyperref}
\usepackage{tabularx}
\usepackage{adjustbox}
\usepackage{xcolor}
\usepackage{nicefrac}
\usepackage{soul}

\setstcolor{red}

\usepackage{hyperref}
\hypersetup{
	colorlinks	=true,
	urlcolor	=blue,
	linkcolor	=blue,
	citecolor	=blue,
	pdftitle	={},
	pdfauthor	={Zhi-Wei Liu, Jun-Xu Lu, Ming-Zhu Liu, Li-Sheng Geng},
	pdfsubject	={$Z_c(3900)$ and $Z_{cs}(3985)$ correlation functions}
	}

\begin{document}

\title{
Femtoscopy can tell whether $Z_c(3900)$ and $Z_{cs}(3985)$ are resonances, virtual states, or bound states
}

\author{Zhi-Wei Liu}
\affiliation{School of Physics, Beihang University, Beijing 102206, China}

\author{Jun-Xu Lu}
\email[Corresponding author: ]{ljxwohool@buaa.edu.cn}
\affiliation{School of Physics, Beihang University, Beijing 102206, China}

\author{Ming-Zhu Liu}
\affiliation{Frontiers Science Center for Rare Isotopes, Lanzhou University,
Lanzhou 730000, China}
\affiliation{ School of Nuclear Science and Technology, Lanzhou University, Lanzhou 730000, China}

\author{Li-Sheng Geng}
\email[Corresponding author: ]{lisheng.geng@buaa.edu.cn}
\affiliation{School of Physics, Beihang University, Beijing 102206, China}
\affiliation{Sino-French Carbon Neutrality Research Center, \'Ecole Centrale de P\'ekin/School of General Engineering, Beihang University, Beijing 100191, China}
\affiliation{Peng Huanwu Collaborative Center for Research and Education, Beihang University, Beijing 100191, China}
\affiliation{Beijing Key Laboratory of Advanced Nuclear Materials and Physics, Beihang University, Beijing 102206, China }
\affiliation{Southern Center for Nuclear-Science Theory (SCNT), Institute of Modern Physics, Chinese Academy of Sciences, Huizhou 516000, China}

\begin{abstract}
There have been extended and heated discussions on the nature of the two exotic states, $Z_c(3900)$ and $Z_{cs}(3985)$, particularly whether they are near-threshold resonances, virtual states, or bound states. In this work, we demonstrate for the first time that the femtoscopic technique can be employed to distinguish between these three scenarios. More concretely, based on the Koonin-Pratt formula with a Gaussian source, we show that the low-momentum $D^0D^{*-}$/$D^0D_s^{*-}$ correlation functions significantly differ in the three scenarios. The high-momentum results exhibit distinct characteristics in the resonant and virtual state scenarios, especially in small collision systems of 1 fm, as produced in $pp$ collisions at the LHC. We hope that these discoveries will stimulate further experimental studies and help clarify the nature of the many exotic states that have been discovered.

\vspace{0.5em}

\noindent Received: 04-May-2025, Revised: 21-Jul-2025, Accepted: 01-Sep-2025
\end{abstract}


\maketitle

\section{Introduction}
In 2013, the BESIII Collaboration and Belle Collaboration observed a hidden-charm tetraquark candidate $Z_c(3900)^{\pm}$ in the $J/\psi\pi^\pm$ invariant mass distribution of the $e^+e^-\rightarrow J/\psi\pi^+\pi^-$ process~\cite{BESIII:2013ris, Belle:2013yex}. This charged charmonium-like state was subsequently confirmed~\cite{Xiao:2013iha, D0:2018wyb}. In 2020, the BESIII Collaboration reported the first signal of a hidden-charm state with strangeness, $Z_{cs}(3985)^{-}$, in the $K^+$ recoil-mass spectrum of the $e^+e^-\rightarrow K^+(D^{*0}D_s^-+D^0D_s^{*-})$ reaction~\cite{BESIII:2020qkh}. Since the quark contents of the charged $Z_c(3900)^{-}/Z_c(3900)^{+}$ and $Z_{cs}(3985)^{-}$ states are $c\bar{c}d\bar{u}$/$c\bar{c}u\bar{d}$ and $c\bar{c}s\bar{u}$, respectively, the latter is often interpreted as the strangeness partner of the former in terms of the SU(3)-flavor symmetry. We note that their isospin quantum numbers have been confirmed by the discovery of their neutral partners~\cite{BESIII:2015cld, BESIII:2022qzr}.

There have been extensive studies on the nature of the $Z_c(3900)$/$Z_{cs}(3985)$ states, treating them as hadronic molecules of $D^0D^{*-}$/$D^0D_s^{*-}$~\cite{Wang:2013cya, Guo:2013sya, Albaladejo:2015lob, He:2017lhy, Guo:2017jvc, Yang:2020nrt, Yan:2021tcp, Meng:2020ihj, Baru:2021ddn, Du:2022jjv, Ji:2022uie, Yan:2023bwt, Chen:2023def}, compact tetraquark states~\cite{Wang:2020iqt, Maiani:2021tri, Wua:2023ntn}, or threshold effects~\cite{Chen:2013coa, Wang:2020kej, Ikeno:2020mra, Prelovsek:2014swa} (see Refs.~\cite{Olsen:2017bmm, Karliner:2017qhf, Yuan:2018inv, Liu:2019zoy, Brambilla:2019esw, Guo:2019twa, Meng:2022ozq} for recent reviews), but no firm conclusion has been reached. Especially due to the lack of reliable theoretical understanding of the $D^0D^{*-}$ and $D^0D_s^{*-}$ interactions, there have been long and heated discussions on whether $Z_c(3900)$ and $Z_{cs}(3985)$ are resonances (poles on the unphysical sheet off the real energy axis), virtual states (poles on the real $s$-axis below the lowest threshold on the unphysical sheet), or bound states (poles on the real $s$-axis below the lowest threshold on the physical sheet). In the experimental analyses based on the Breit-Wigner parametrization, $Z_c(3900)$ and $Z_{cs}(3985)$ are suggested to be typical resonances~\cite{BESIII:2013ris, Belle:2013yex, BESIII:2020qkh}. However, various theoretical re-analyses of the same data suggest they can be either resonances, virtual states, or bound states~\cite{Guo:2013sya, Albaladejo:2015lob, He:2017lhy, Guo:2017jvc, Yang:2020nrt, Yan:2021tcp, Meng:2020ihj, Baru:2021ddn, Du:2022jjv, Ji:2022uie, Yan:2023bwt, Chen:2023def}. The lattice QCD simulations of coupled-channel $\pi J/\psi$, $\rho \eta_c$, and $\bar{D}D^*$ interactions at $m_\pi=400-710$ MeV even indicates that the $Z_c(3900)$ is a threshold cusp~\cite{HALQCD:2016ofq}, which can produce the lineshapes of $Y(4260)$ consistent with the BESIII data~\cite{BESIII:2013ris, BESIII:2013qmu}.

In recent years, femtoscopy, which measures two-hadron momentum correlation functions (CFs), has made remarkable progress in revealing the dynamics of the strong interactions between pairs of stable and unstable hadrons~\cite{ExHIC:2017smd, Fabbietti:2020bfg, Liu:2024uxn}. Thanks to the abundant rare hadrons produced and the various particle emission sources available in relativistic $pp$, $pA$, and $AA$ collisions, the femtoscopic technique can be used not only to study the strong interactions involving unstable hadrons that are difficult or impossible in traditional scattering experiments but also to provide more and complementary information compared with the measurement of invariant mass distributions. A lot has been learned about the meson-meson~\cite{ALICE:2017jto, ALICE:2018nnl, ALICE:2020mkb, ALICE:2021ovd, ALICE:2021ovd, ALICE:2023eyl}, meson-baryon~\cite{ALICE:2019gcn, ALICE:2020wvi, ALICE:2021cpv, ALICE:2022yyh, ALICE:2023wjz}, and (anti)baryon-(anti)baryon ~\cite{STAR:2014dcy, STAR:2015kha, STAR:2018uho, Isshiki:2021bqh, ALICE:2018ysd, ALICE:2018ysd, ALICE:2018ysd, ALICE:2019eol, ALICE:2019buq, ALICE:2019igo, ALICE:2019hdt, ALICE:2020mfd, ALICE:2021njx, ALICE:2021cyj, ALICE:2021njx, ALICE:2022uso} interactions in the light quark (u, d, s) sector with the femtoscopic technique. More recently, the successful measurements of the $\bar{D}N$, $D^{(*)}\pi$, and $D^{(*)}K$ CFs demonstrated the huge potential to access the charm sector experimentally~\cite{ALICE:2022enj, ALICE:2024bhk}. With the upgraded ALICE apparatus and the larger data sample expected at LHC runs 3, 4, and 5, more reactions can be accessed with high precision in the future~\cite{ALICE:2022wwr}.

On the other hand, the experimental studies have also triggered a large number of theoretical studies~\cite{Morita:2014kza, Morita:2016auo, Ohnishi:2016elb, Haidenbauer:2018jvl, Morita:2019rph, Kamiya:2019uiw, Ogata:2021mbo, Kamiya:2021hdb, Haidenbauer:2021zvr, Liu:2022nec, Molina:2023jov, Molina:2023oeu, Sarti:2023wlg}, especially in the charm and bottom sectors~\cite{Haidenbauer:2020kwo, Kamiya:2022thy, Liu:2023uly, Liu:2023wfo, Vidana:2023olz, Ikeno:2023ojl, Albaladejo:2023pzq, Torres-Rincon:2023qll, Albaladejo:2023wmv, Feijoo:2023sfe, Khemchandani:2023xup, Li:2024tof, Feijoo:2024bvn}. It was shown that femtoscopy offers a valuable means to directly test the hadronic molecular picture of the exotic hadrons~\cite{Liu:2023uly, Liu:2023wfo, Vidana:2023olz, Ikeno:2023ojl}. However, it is worth noting that the above studies primarily focus on bound states, with limited discussion of resonant or virtual states. In this letter, we demonstrate how the femtoscopic technique can be utilized to study hadronic interactions in the presence of near-threshold resonances, virtual states, or bound states. In particular, assuming that $Z_c(3900)$/$Z_{cs}(3985)$ is a molecular state of $D^0D^{*-}$/$D^0D_s^{*-}$, the femtoscopic technique shows great potential in distinguishing whether $Z_c(3900)$/$Z_{cs}(3985)$ is a resonance, a virtual state, or a bound state.

\section{Theoretical framework}
Close to the threshold, the most general $S$-wave potential up to the next-to-leading order in momentum expansion has the following form
\begin{align}\label{Eq:V}
  V(k)=a+b~k^2,
\end{align}
where $k=\frac{\sqrt{s-(m_1+m_2)^2}\sqrt{s-(m_1-m_2)^2}}{2\sqrt{s}}$ represents the center-of-mass (c.m.) momentum of the particle pair with masses $m_{1,2}$, and $a$ and $b$ are (real) low-energy constants (LECs) to be determined. Then, the single-channel $T$-matrix can be obtained as
\begin{align}
  T(\sqrt{s})&=\left(1-VG\right)^{-1}V=-\rho^{-1}e^{i\delta}\sin\delta\label{Eq:Tmatrix2},
\end{align}
where $\sqrt{s}$ represents the c.m. energy, $\delta$ is the phase shift, and $\rho=k/\left(8\pi\sqrt{s}\right)$ is the phase-space factor. 
The loop function (in the first/physical Riemann sheet (RS)) $G$ is regularized by a sharp cutoff $q_{\rm max}$ of the order of $0.6-1.4$ GeV,
\begin{align}
  G(\sqrt{s})=\int\limits_0^{q_{\rm max}}\frac{{\rm d}^3k'}{(2\pi)^3}\frac{E_1'+E_2'}{2E_1'E_2'}\frac{1}{s-(E_1'+E_2')^2+i\varepsilon},\label{Eq:LF}
\end{align}
with $E_i'=\sqrt{m_i^2+k^{\prime2}}$. Here, virtual (bound) and resonant states are identified as poles in the second (first) RS of the $T$-matrix on the real axis below the threshold and the complex $\sqrt{s}$ plane, respectively. For a resonance, the real and imaginary parts of the pole position define its mass and half-width. The loop function in the second/unphysical RS, $G^{\rm II}$, is related to that in the first RS by an analytic continuation as $G^{\rm II}=G+2i\rho$~\cite{Oller:1997ti}. In addition, it is worth calculating the scattering length $a_0$ and effective range $r_{\rm eff}$~\cite{Ikeno:2023ojl}
\begin{subequations}
\begin{align}
  -\frac{1}{a_0}&=\left[-8\pi\sqrt{s}T^{-1}\right]_{m_1+m_2},\label{Eq:a0}\\
  r_{\rm eff}&=\left[\frac{\partial^2}{\partial k^2}\left(-8\pi\sqrt{s}T^{-1}+ik\right)\right]_{m_1+m_2}.\label{Eq:reff}
\end{align}
\end{subequations}

\begin{table*}[htpt]
\centering
\setlength{\tabcolsep}{10.6pt}
\caption{Masses and widths of $Z_c(3900)$ and $Z_{cs}(3985)$, the LECs $a$, $b$, the scattering length $a_0$, and the effective range $r_{\rm eff}$.\footnote{The $D^0D^{*-}$ and $D^0D_s^{*-}$ thresholds are $3875.1$ MeV and $3977.04$ MeV, respectively. The central values of $a$, $b$, $a_0$, and $r_{\rm eff}$ are obtained with $q_{\rm max}$$=1.0$ GeV. The upper and lower deviations correspond to the calculations of $q_{\rm max}=0.6$ and $1.4$ GeV, respectively.}}\label{Tab:T2}
\begin{tabular}{cccccccc}
\hline
                  & Scenario  & \multicolumn{1}{c}{$M$~(MeV)}  & \multicolumn{1}{c}{$\Gamma$~(MeV)}  & \multicolumn{1}{c}{$a$}  & \multicolumn{1}{c}{$b$~($10^{-6}\cdot$ MeV$^{-2}$)}  & \multicolumn{1}{c}{$a_0$~(fm)}  & \multicolumn{1}{c}{$r_{\rm eff}$~(fm)}  \\ \hline
\multirow{3}{*}{$Z_c(3900)$} & Resonant  & $3887.1$~\cite{ParticleDataGroup:2024cfk}  & $28.4$~\cite{ParticleDataGroup:2024cfk}  & $-101.7_{-9.6}^{+15.7}$  & $-1380.6_{-1800.4}^{+586.8}$  & $-0.6_{+0.2}^{-0.1}$  & $-4.8_{-4.6}^{+0.9}$  \\
                  & Virtual  & $3871$  & $0$  & $-137.9_{-67.5}^{+32.2}$  & $0$  & $-2.1_{+0.1}^{-0.0}$  & $0.3_{+0.2}^{-0.0}$  \\
                  & Bound  & $3867$  & $0$  & $-196.0_{-167.6}^{+59.0}$  & $0$  & $1.7_{+0.1}^{-0.0}$  & $0.3_{+0.2}^{-0.1}$  \\ \hline
\multirow{3}{*}{$Z_{cs}(3985)$} & Resonant  & $3988$~\cite{ParticleDataGroup:2024cfk}  & $14$~\cite{ParticleDataGroup:2024cfk}  & $-87.9_{+17.3}^{+8.7}$  & $-2686.7_{-3866.3}^{+1178.8}$  & $-0.4_{+0.2}^{-0.1}$  & $-13.4_{-38.0}^{+4.2}$  \\
                  & Virtual  & $3974$  & $0$  & $-144.0_{-73.0}^{+34.3}$  & $0$  & $-2.4_{+0.1}^{-0.0}$  & $0.3_{+0.2}^{-0.1}$  \\
                  & Bound  & $3972$  & $0$  & $-192.5_{-155.4}^{+56.6}$  & $0$  & $2.1_{+0.1}^{-0.0}$  & $0.3_{+0.2}^{-0.1}$  \\
\hline
\end{tabular}
\end{table*}

\begin{figure*}[htbp]
  \centering
  \includegraphics[width=1\textwidth]{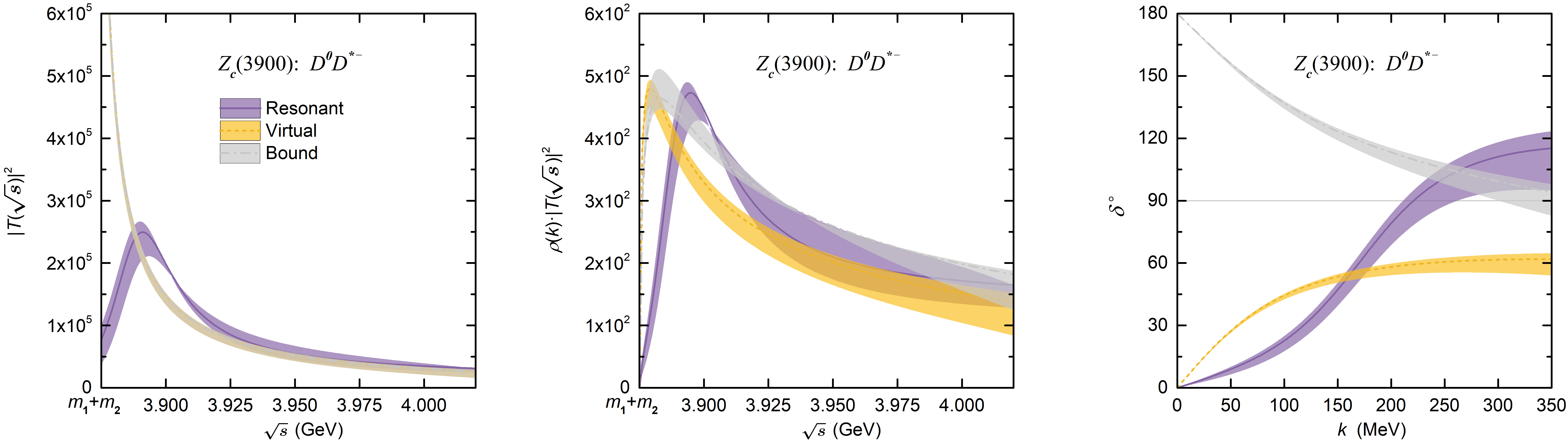}
  \includegraphics[width=1\textwidth]{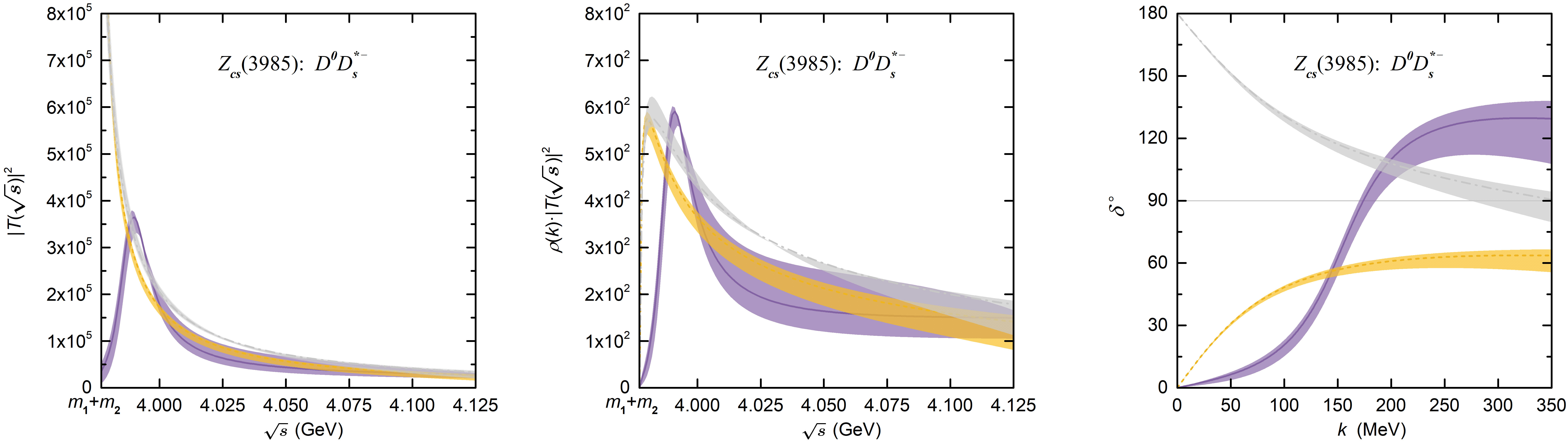}
  \caption{$D^0D^{*-}$ (upper) and $D^0D_s^{*-}$ (lower) $|T|^2$, $\rho|T|^2$ as a function of the c.m. energy $\sqrt{s}$, and the phase shifts as a function of the relative momentum $k$. The bands reflect the variation of the sharp cutoff in the range of $q_{\rm max}=0.6 - 1.4$ GeV.}\label{Fig:Tmatrix_new}
\end{figure*}

According to the Koonin-Pratt formula~\cite{Koonin1977PLB70.43, Pratt1990PRC42.2646, Bauer1992ARNPS42.77}, the femtoscopic CF depends on two quantities, namely, $\psi$ the scattering wave function of the relative motion for the pair of interest and $S_{12}$, the particle-emitting source created in relativistic $pp$, $pA$, and $AA$ collisions. The former contains the information on the final-state interaction we are interested in. In general, there are two ways to obtain the scattering wave function, either by solving the Schr\"odinger equation in coordinate space~\cite{ALICE2018EPJC78.394, Kamiya:2021hdb} or the Lippmann-Schwinger (Bethe-Salpeter) equation in momentum space~\cite{Haidenbauer:2018jvl, Liu:2023uly}. In this work, since the potential of Eq.~\eqref{Eq:V} is constructed in momentum space, it is convenient first to obtain the reaction amplitude $T$-matrix by solving Eq.~\eqref{Eq:Tmatrix2} and then derive the scattering wave function using the relation $|\psi\rangle=|\varphi\rangle+GT|\varphi\rangle$, where $G$ and $|\varphi\rangle$ represent the free propagator and wave function. For the particle-emitting source, a data-driven approach named the resonance source model proposed by the ALICE Collaboration has been widely employed in the femtoscopic analyses~\cite{ALICE:2021njx, ALICE:2019buq, ALICE:2020mfd, ALICE:2021cpv, ALICE:2021cyj, ALICE:2022enj},  based on the hypothesis of a common emission source for all hadrons in small collision systems. In this model, the source is characterized by a Gaussian core that emits all primordial particles, complemented by an exponential tail arising from strongly decaying resonances~\cite{ALICE:2020ibs, ALICE:2023sjd}. The core size $R$ shows a clear transverse mass ($m_{\rm T}$) scaling~\cite{ALICE:2020ibs, ALICE:2023sjd}, which is commonly associated with collective effects (such as radial flow as previously observed in heavy-ion collisions) and has been reproduced successfully in the single-particle emission model ``\textit{Common Emission in CATS}'' (CECA) by introducing spacial-momentum correlations~\cite{Mihaylov:2023pyl}. The resonance yields and decay kinematics are estimated using the thermal and transport models~\cite{Andronic:2017pug, Pierog:2013ria}. For a fixed number of particle pairs, an increase in the number of large-distance (the exponential tail) pairs leads to a larger average pair separation. Consequently, replacing the single Gaussian source with a source with an exponential tail will lead to a weakening of the computed CF strength. In the present work, given the expected minimal contribution from the strong decays feeding into charm mesons~\cite{ALICE:2022enj}, it is pertinent to adopt a Gaussian source $S_{12}=\exp\left[-r^2/(4R^2)\right]/(2\sqrt{\pi}R)^3$ in this exploratory study. The source size $R$ can be anchored to the proton-proton correlation data by measuring $m_{\rm T}$ of the pair of interest in future experiments. With the scattering amplitude obtained directly from Eq.~\eqref{Eq:Tmatrix2} and the Gaussian source, the CF is expressed as~\cite{Vidana:2023olz}
\begin{subequations}
\begin{align}
  C(k)&=1+\int\limits_0^\infty{\rm d}^3r~S_{12}\left[\left|j_0+T\widetilde{G}\right|^2-|j_0|^2\right],\label{Eq:CF_Tmatrix}\\
  &=1+\mathcal{F}_1\sin^2\delta+\mathcal{F}_2\sin\delta\cos\delta,\label{Eq:CF_PS}
\end{align}
\end{subequations}
where $j_0(kr)$ is the spherical Bessel function. The functions $\mathcal{F}_1$ and $\mathcal{F}_2$ are given by
\begin{subequations}
\begin{align}
  \mathcal{F}_1(R,k)&=\int\limits_0^\infty{\rm d}^3r~S_{12}\left[\left(\frac{{\rm Re}\widetilde{G}}{\rho}\right)^2-\left|j_0\right|^2\right],\label{Eq:F1_1}\\
  \mathcal{F}_2(R,k)&=-\int\limits_0^\infty{\rm d}^3r~S_{12}\left(2j_0\cdot\frac{{\rm Re}\widetilde{G}}{\rho}\right),\label{Eq:F2_1}
\end{align}
\end{subequations}
which are divergent at $k=0$. The quantity $\widetilde{G}$ is given by
\begin{align}
  \widetilde{G}(r,\sqrt{s})=\int\limits_0^{q_{\rm max}}&\frac{{\rm d}^3k'}{(2\pi)^3}\frac{E_1'+E_2'}{2E_1'E_2'}\frac{j_0(k'r)}{s-(E_1'+E_2')^2+i\varepsilon}.\label{Eq:LFj0}
\end{align}

A recent work~\cite{Epelbaum:2025aan} pointed out that the off-shell ambiguity of the strong interaction leads to changes in the short-distance behavior of the wave function, thereby introducing uncertainty in the calculation of correlation functions. However, two subsequent papers~\cite{Gobel:2025afq, Molina:2025lzw} demonstrated that the uncertainty caused by the off-shell effect may be relatively mild when using ``realistic'' strong interactions. In the present work, we adopt the same starting point as in Ref.~\cite{Vidana:2023olz}, namely, the derivation of Eq.~\eqref{Eq:CF_Tmatrix} starts from a separable potential $V\cdot\theta(q_{\rm max}-k)\cdot\theta(q_{\rm max}-k')$ in momentum space, which reverts into the separable scattering matrix $T\cdot\theta(q_{\rm max}-k)\cdot\theta(q_{\rm max}-k')$. The on-shell $T$ and $\theta(q_{\rm max}-k)$ factor are factored out of the integral, and $\theta(q_{\rm max}-k)$ factor is inoperative since one always works with values of the momenta smaller than $q_{\rm amx}$. The $\theta(q_{\rm max}-k')$ factor, which controls the off-shell behavior, goes into the integral of $\widetilde{G}$ and is absorbed as the upper limit of the integral. For more details, see Refs.~\cite{Vidana:2023olz, Molina:2025lzw}. Therefore, varying the sharp cutoff $q_{\rm max}$ provides a way to assess the uncertainty associated with off-shell effects.

\begin{figure*}[htbp]
  \centering
  \includegraphics[width=1\textwidth]{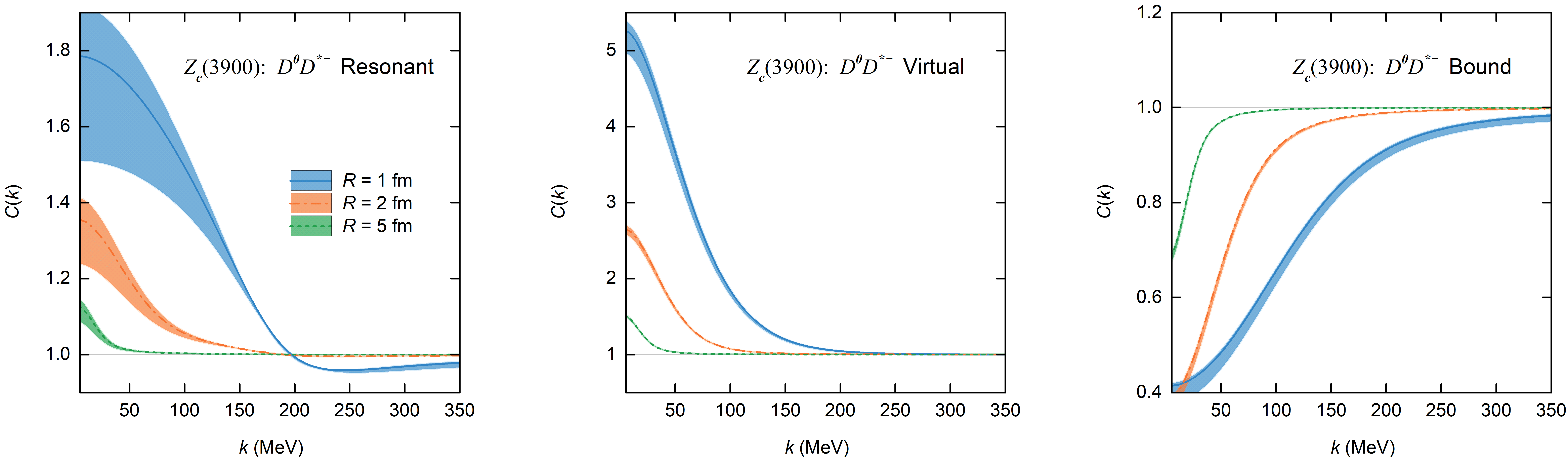}
  \includegraphics[width=1\textwidth]{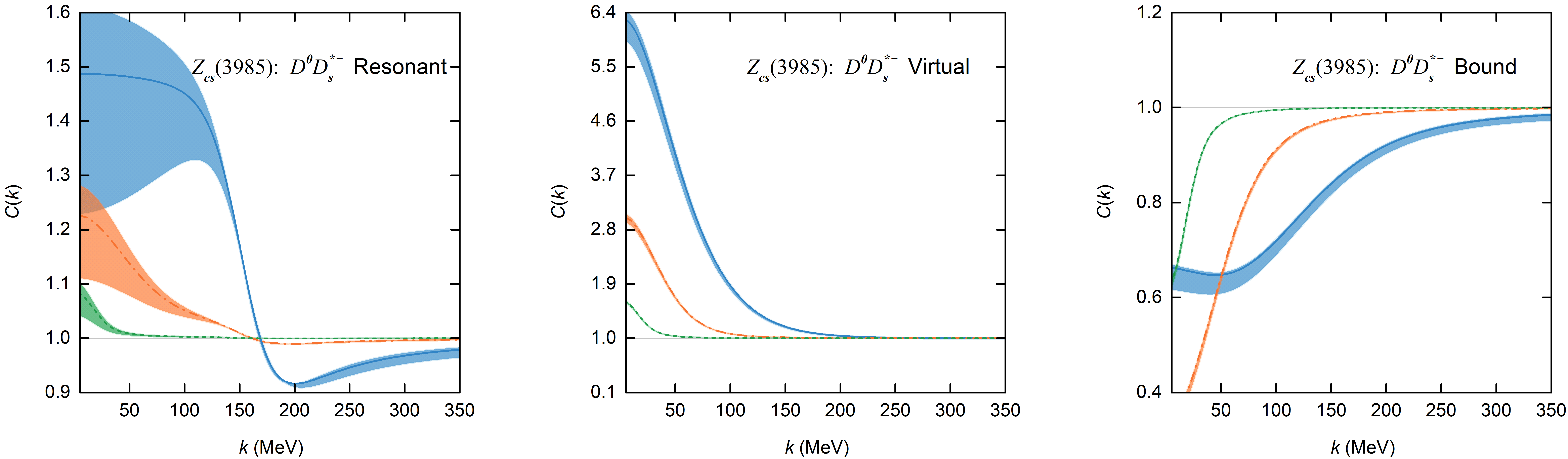}
  \caption{$D^0D^{*-}$ (upper) and $D^0D_s^{*-}$ (lower) correlation functions as a function of the relative momentum $k$ for different source sizes $R = 1$, $2$, and $5$ fm. The bands reflect the variation of the sharp cutoff in the range of $q_{\rm max}=0.6 - 1.4$ GeV.}\label{Fig:CF_new}
\end{figure*}

As mentioned above, $a$ and $b$ are the only independent parameters in describing the $D^0D^{*-}$/$D^0D_s^{*-}$ interaction. They can be determined by reproducing the pole positions of $Z_c(3900)$/$Z_{cs}(3985)$ in the three scenarios of resonant/virtual/bound states. We adopt the latest PDG averages for the resonant state poles~\cite{ParticleDataGroup:2024cfk} and fine-tune the virtual/bound state pole to obtain a line shape for the $D^0D^{*-}$/$D^0D_s^{*-}$ invariant mass spectrum (IMS) similar to that of the resonant state scenario. These pole settings, LECs $a$, $b$, the scattering length, and the effective range are listed in Table~\ref{Tab:T2}. The adopted virtual state poles of $Z_c(3900)$ and $Z_{cs}(3985)$ are compatible with the theoretical results~\cite{Guo:2013sya, Albaladejo:2015lob, He:2017lhy, Yang:2020nrt, Baru:2021ddn, Du:2022jjv, Ji:2022uie, Chen:2023def}, and the adopted bound state poles of $Z_c(3900)$ and $Z_{cs}(3985)$ are close to the results of Ref.~\cite{Baru:2021ddn}. In addition, it has been demonstrated that virtual and bound states generated by a constant potential are predominantly hadronic molecules~\cite{Guo:2017jvc}. However, introducing an additional momentum-dependent term can lead to the emergence of non-molecular components. The large negative effective range, as listed in Table~\ref{Tab:T2}, suggests that the non-molecular component may play a relevant role in the resonant state scenario of $Z_c(3900)$/$Z_{cs}(3985)$.

\section{Results and discussion}
Before studying CFs, it is necessary to discuss the properties of $|T|^2$, $\rho|T|^2$, and $\delta$ for the three scenarios, as shown in Fig.~\ref{Fig:Tmatrix_new}. Since the virtual and bound state poles lie on the real axis below the threshold, while the resonant state pole lies above the threshold and deviates from the real axis, $|T|^2$ for virtual and bound states near the threshold $m_1 + m_2$ are significantly larger than those for resonant states. However, $|T|^2$ is not directly observable; the measurable IMS is proportional to $\rho|T|^2$~\cite{Dong:2020nwy}. Due to the phase-space suppression near the threshold, IMSs of virtual and bound states exhibit resonance-like shapes, as $\rho|T|^2$ depicts. This explains why it is difficult to distinguish near-threshold resonant/virtual/bound states through IMSs with limited experimental resolution. On the other hand, $\delta$ exhibits unique features in the three scenarios. For instance, the resonant and virtual states $\delta$ start from $0^{\circ}$ and increase with momentum/energy; the bound state $\delta$ starts at $180^{\circ}$ and decreases; as momentum increases, the resonant state $\delta$ exceeds $90^{\circ}$ around the resonant energy.

We present the $D^0D^{*-}$ (upper) and $D^0D_s^{*-}$ CFs (lower) for three typical source sizes $R = 1$, $2$, and $5$ fm in Fig.~\ref{Fig:CF_new}, where the left/middle/right panel corresponds to the results in the resonant/virtual/bound state scenarios. In the resonant state scenario, the $D^0D^{*-}$ CF is enhanced (compared to unity) in the low-momentum region but suppressed in the high-momentum region and exhibits a minimum. This resonance behavior can be understood from Eq.~\eqref{Eq:CF_PS}: as the momentum increases and $\delta$ crosses $90^{\circ}$, $\mathcal{F}_2\sin\delta\cos\delta$ changes from positive to negative because $\mathcal{F}_2$ is always positive in the present work. In the virtual state scenario, the $D^0D^{*-}$ CF shows an enhancement in a wide range of $k$, with a particularly strong enhancement in the low-momentum region, which can be traced to the increasing virtual state $|T|^2$ near the threshold. In the bound state scenario, the $D^0D^{*-}$ CF is suppressed in a wide range of $k$, which is similar to the result obtained with a strongly attractive potential in the square-well model~\cite{Liu:2023uly} (this point can also be obtained by analyzing the bound state $\delta$). As the source size $R$ decreases, the difference in the CFs between these three scenarios becomes more significant because of the short-range nature of the strong interaction.

Due to the SU(3)-flavor symmetry, the $D^0D_s^{*-}$ CFs exhibit similar behaviors as the $D^0D^{*-}$ CFs. However, in the resonant state scenario, the suppression in the high-momentum region becomes more pronounced, which can be attributed to the narrower width of $Z_{cs}(3985)$. This is understandable because as the pole approaches the real axis, the influence of the resonant structure on physical observables intensifies. We emphasize that experimentally it is feasible to distinguish whether $Z_c(3900)$/$Z_{cs}(3985)$ is a resonant/virtual/bound state by measuring the $D^0D^{*-}$/$D^0D_s^{*-}$ CFs, especially in $pp$ collisions. This is because, on the one hand, the three CFs differ significantly in numerical values and exhibit distinct characteristics in the low-momentum region. On the other hand, the experimental uncertainties in the high-momentum region are usually much smaller than those in the low-momentum region (see, e.g., Refs~\cite{STAR:2014dcy, STAR:2015kha, ALICE:2017jto, ALICE:2018nnl, ALICE:2018ysd, STAR:2018uho, ALICE:2019buq, ALICE:2019eol, ALICE:2019gcn, ALICE:2019hdt, ALICE:2019igo, ALICE:2020mfd, ALICE:2020mkb, ALICE:2020wvi, ALICE:2021cpv, ALICE:2021cyj, ALICE:2021njx, ALICE:2021ovd, Isshiki:2021bqh, ALICE:2022enj, ALICE:2022uso, ALICE:2022yyh, ALICE:2023eyl, ALICE:2023wjz, ALICE:2024bhk}). Our results show that for the $D^0D^{*-}$ and $D^0D_s^{*-}$ CFs, the relative difference in the CFs between the resonant and virtual scenarios can be as significant as about $7\%$ and $17\%$ for $R = 1$ fm, and even about $10\%$ and $23\%$ for $R = 0.9$ fm at $k\approx230$ and $k\approx180$ MeV, respectively.

Before closing our discussion, some comments are relevant in understanding the robustness of the results: i) Coupled-channel effects~\cite{Haidenbauer:2018jvl, Feijoo:2024bvn}. For the $D^0D^{*-}$ CF, the $\pi J/\psi-D\bar{D}^*$ coupled-channel effect may offset the high-momentum suppression in the resonant state scenario. However, the significant differences in the low-momentum region can still distinguish the three scenarios. For the $D^0D_s^{*-}$ CF, the $\bar{K}J/\psi-D\bar{D}_s^*$ coupled-channel effect can be safely neglected due to the small inelastic transition between $D\bar{D}_s^*$ and $\bar{K}J/\psi$~\cite{BESIII:2023wqy}. ii) Effect of the uncertainty of pole positions. We have checked that even accounting for the uncertainties of pole positions (e.g., $Z_{cs}(3985)$ resonant pole $3988^{+5}_{-5}+i7^{+2}_{-2}$ MeV~\cite{ParticleDataGroup:2024cfk}, virtual pole $3974^{+1.3}_{-1.0}$ MeV, and bound pole $3972^{+2.5}_{-2.2}$ MeV), CFs can still distinguish whether $Z_c(3900)$ and $Z_{cs}(3985)$ are resonant, virtual, or bound states. In addition, stricter constraints, especially for the shallow bound state, can be obtained by measuring the CFs in $pp$, $pA$, and $AA$ collisions. iii) Comparison with the Lednick${\rm \acute{y}}$-Lyuboshits (LL) model~\cite{Lednicky:1981su, ExHIC:2017smd, Fabbietti:2020bfg}. With the scattering parameters listed in Table~\ref{Tab:T2}, the LL model can reproduce the CFs for virtual and bound state scenarios. However, for resonant state scenarios where $|r_{\rm eff}|$ is much larger than the source size $R$, the effective range correction may introduce significant deviations. (see the Supplemental Material for more details.) In addition, it is worthwhile to note that the recent experimental observations of a marked enhancement in low-momentum neutron-neutron correlation functions provide strong experimental validation for our predictions regarding virtual state correlations~\cite{Si:2025eou}.

\section{Summary and outlook}
In this work, we proposed to decipher the nature of $Z_c(3900)$ and $Z_{cs}(3985)$ with the femtoscopic technique. We first evaluated the $D^0D^{*-}$/$D^0D_s^{*-}$ interactions by reproducing the pole positions of $Z_c(3900)$/$Z_{cs}(3985)$ in the resonance, virtual state, and bound state scenarios, respectively. We noted that the phase-space suppression effect is the main reason for the difficulty in distinguishing the near-threshold states using the invariant mass spectrum alone. Based on the Koonin-Pratt formula with a Gaussian source, we found that the low-momentum $D^0D^{*-}$/$D^0D_s^{*-}$ correlation functions are significantly different in the three scenarios. The high-momentum results exhibit distinct characteristics in the resonant and virtual state scenarios, especially in small collision systems of the order of 1 fm, as produced in $pp$ collisions at the LHC. The novel approach can help distinguish whether $Z_c(3900)$ and $Z_{cs}(3985)$ are resonances, virtual states, or bound states , which is crucial to confirm whether $Z_c(3900)$ and $Z_{cs}(3985)$ observed by BESIII and LHCb are genuine resonances or not. In addition, we derived a correspondence between the femtoscopic correlation function and the scattering phase shifts, which can help extract the strong interactions between hadron pairs from correlation functions.

Considering the experimental precision achievable at present and future facilities, we strongly encourage measuring the $D^0D^{*-}$/$D^0D_s^{*-}$ correlation functions in $pp$, $pA$, and $AA$ collisions. Given that most of the exotic hadronic states discovered experimentally are located near the thresholds of two ordinary hadron pairs, this work explores a unified framework to distinguish near-threshold resonances, virtual states, or bound states in these hadron pairs, especially for systems without the Coulomb interaction and with negligible coupled-channel effects. In fact, the closer to the threshold, the more difficult it is to identify these three states through invariant mass spectra. Conversely, it may be easier to identify the nature of these states through the femtoscopic correlation function. Therefore, it has great potential for understanding exotic hadrons by combining studies of correlation functions and invariant mass spectra.

\section*{Conflict of interest}
The authors declare that they have no conflict of interest.

\section*{Acknowledgments}
We thank Eulogio Oset for a careful reading of the original manuscript. We also thank the anonymous referees for their valuable comments, which helped greatly improve our manuscript. This work is partly supported by the National Key R\&D Program of China (2023YFA1606703) and the National Natural Science Foundation of China (12435007). Zhi-Wei Liu acknowledges support from the National Natural Science Foundation of China (12405133, 12347180), China Postdoctoral Science Foundation (2023M740189), and the Postdoctoral Fellowship Program of CPSF (GZC20233381).

\section*{Author contributions}
Zhi-Wei Liu and Li-Sheng Geng conceived the work and prepared the manuscript. All the authors contributed to the theoretical derivation, numerical calculation, and discussion of the results. All the authors have read and approved the final version of the manuscript.

\bibliographystyle{elsarticle-num}
\bibliography{Zc}

\section*{Supplemental Material}
In this Supplemental Material, we provide further information that substantiates the results and conclusion obtained in the main text, including coupled-channel effects, input uncertainties, and the performance of the Lednick${\rm \acute{y}}$-Lyuboshits model. 
\subsection{Coupled-channel effects}
In this section, in addition to the $D\bar{D}^*$ channel, we consider the relevant channel of  $\pi J/\psi$, which plays an important role in the decay of $Z_c(3900)$~\footnote{In addition to the $D\bar{D}^*$ and $\pi J/\psi$ channels, there are $\rho\eta_c$, $\pi\psi(2S)$ and $D^*\bar{D}^*$ channels. For the energy below the $D^*\bar{D}^*$ threshold, the corresponding $D^*\bar{D}^*\rightarrow D\bar{D}^*$ wave functions drop exponentially, leading to suppressed $D^*\bar{D}^*$ contributions in the $D\bar{D}^*$ CF. Furthermore, the contributions from the OZI-suppressed $\rho\eta_c$ and $\pi\psi(2S)$ coupled channels can be effectively absorbed into the contact term of the inelastic $D\bar{D}^*\rightarrow\pi J/\psi$ transition. Therefore, to reduce the number of free parameters, we only consider $D\bar{D}^*$ and $\pi J/\psi$ channels in this work.}. We refer to the $\pi J/\psi$ as channel $1$ and $D\bar{D}^*$ as channel $2$. The coupled-channel potential reads
\begin{align}
  V_{\rm 2ch}(\sqrt{s})&=
  \left(
  \begin{array}{cc}
    0 & c+d~\frac{k_1^2+k_2^2}{2} \\
    c+d~\frac{k_1^2+k_2^2}{2} & a+b~k_2^2
  \end{array}
  \right),\label{Eq:V2ch}
\end{align}
where $k_i=\frac{\sqrt{s-(m_{i1}+m_{i2})^2}\sqrt{s-(m_{i1}-m_{i2})^2}}{2\sqrt{s}}$ represents the c.m. momentum of the $i$th channel, $a$, $b$, $c$, and $d$ are real LECs to be determined. The $\pi J/\psi\rightarrow\pi J/\psi$ interaction is neglected due to the Okubo-Zweig-Iizuka (OZI) suppression. Then, the coupled-channel $T$-matrix can be obtained as
\begin{align}
  T_{\rm 2ch}(\sqrt{s})=\left[\mathbb{I}-V_{\rm 2ch}G_{\rm 2ch}\right]^{-1}V_{\rm 2ch},\label{Eq:T2ch}
\end{align}
where $G_{\rm 2ch}$ is the diagonal matrix of loop functions. The generalized coupled-channel CF for a specific channel $j$ reads~\cite{Haidenbauer:2018jvl, Vidana:2023olz}
\begin{align}
  C_j(k_j)&=1+\int\limits_0^\infty{\rm d}^3r~S_{12}~\bigg[\sum_i\omega_i\left|\delta_{ij}j_0(k_jr)+T_{ij}\widetilde{G}_i\right|^2\nonumber\\
  &-\left|j_0(k_jr)\right|^2\bigg],\label{Eq:CF2ch}
\end{align}
where $\omega_i$ is the weight for the individual components of the multi-channel wave function. Due to the lack of experimental information, we use the simplest statistical model~\cite{Andronic:2005yp, Kamiya:2019uiw} to estimate the ratio of the weight $\omega_i/\omega_j$, which reads
\begin{align}
  \frac{\omega_i}{\omega_j}\approx\frac{\exp[(-m_{i1}-m_{i2})/T^*]}{\exp[(-m_{j1}-m_{j2})/T^*]},\label{Eq:weight}
\end{align}
where the hadronization temperature is $T^*=154$ MeV~\cite{HotQCD:2014kol, Borsanyi:2013bia}. The weight of the observed channel $\omega_{D^0D^{*-}}$ is set at $1$, and the calculated $\omega_{\pi^-J/\psi}$ equals $63.2$ ($\omega_{K^-J/\psi}/\omega_{D^0D_s^{*-}}\approx12.3$ for $Z_{cs}(3985)$).

\begin{table*}[htpt]
\centering
\setlength{\tabcolsep}{10.6pt}
\caption{Masses and widths of $Z_c(3900)$, and the LECs $a$, $b$, $c$, and $d$ in the coupled-channel framework.\footnote{The sharp cutoff $q_{\rm max}$ is fixed at $1$ GeV.}}\label{Tab:T3}
\begin{tabular}{cccccccc}
\hline
                  & Scenario  & \multicolumn{1}{c}{$M$~(MeV)}  & \multicolumn{1}{c}{$\Gamma$~(MeV)}  & \multicolumn{1}{c}{$a$}  & \multicolumn{1}{c}{$b$~($10^{-6}\cdot$ MeV$^{-2}$)}  & \multicolumn{1}{c}{$c$}  & \multicolumn{1}{c}{$d$~($10^{-6}\cdot$ MeV$^{-2}$)}  \\ \hline
                  & Resonant  & $3887.1$~\cite{ParticleDataGroup:2024cfk}  & $28.4$~\cite{ParticleDataGroup:2024cfk}  & $-99.0$  & $-1594.2$  & $-123.6$  & $374.4$  \\
     $Z_c(3900)$  & Virtual  & $3871.7$  & $2.0$  & $-138.9$  & $0$  & $-19.4$  & $0$  \\
                  & Bound  & $3868.4$  & $0.9$  & $-192.0$  & $0$  & $-13.2$  & $0$  \\
\hline
\end{tabular}
\end{table*}

For the resonant state scenario, we utilize the mass and width of $Z_c(3900)$ suggested by the PDG~\cite{ParticleDataGroup:2024cfk}. Due to the introduction of new free parameters $c$ and $d$, we further consider the experimental branching ratio $BR=\frac{\Gamma(Z_c(3885)\rightarrow D\bar{D}^*)}{\Gamma(Z_c(3900)\rightarrow\pi J/\psi)}=6.2$~\cite{BESIII:2013qmu} ($\frac{\Gamma(Z_{cs}(3985)^+\rightarrow K^+J/\psi)}{\Gamma(Z_{cs}(3985)^+\rightarrow (\bar{D}^0D_s^{*+}+\bar{D}^{*0}D_s^+)}<0.03$~\cite{BESIII:2023wqy}). Since the $D\bar{D}^*$ threshold is about $640$ MeV above the $\pi J/\psi$ threshold, it is expected that the introduction of the $\pi J/\psi$ channel mainly affects the $Z_c(3900)$ width rather than its mass. Therefore, we first switch off the $\pi J/\psi$ channel and use the $Z_c(3900)$ mass and partial width ($\Gamma_{D\bar{D}^*} = \Gamma\frac{BR}{BR+1}\approx24.5$ MeV) to fix the LECs $a$ and $b$ (similar to the single-channel study of the main text), and then use the mass and the full width ($\Gamma = 28.4$ MeV) to determine the remaining LECs $c$ and $d$ in the coupled-channel framework. For the virtual and bound state scenarios, only the constants $a$ and $c$ are retained in the potential, while $b$ and $d$ are set to $0$. The corresponding poles are fine-tuned to obtain a line shape for the $D^0D^{*-}$ IMS similar to the resonant state scenario. The pole settings, and the LECs $a$, $b$, $c$, and $d$ are listed in Table~\ref{Tab:T3}.

\begin{figure*}[htbp]
  \centering
  \includegraphics[width=1\textwidth]{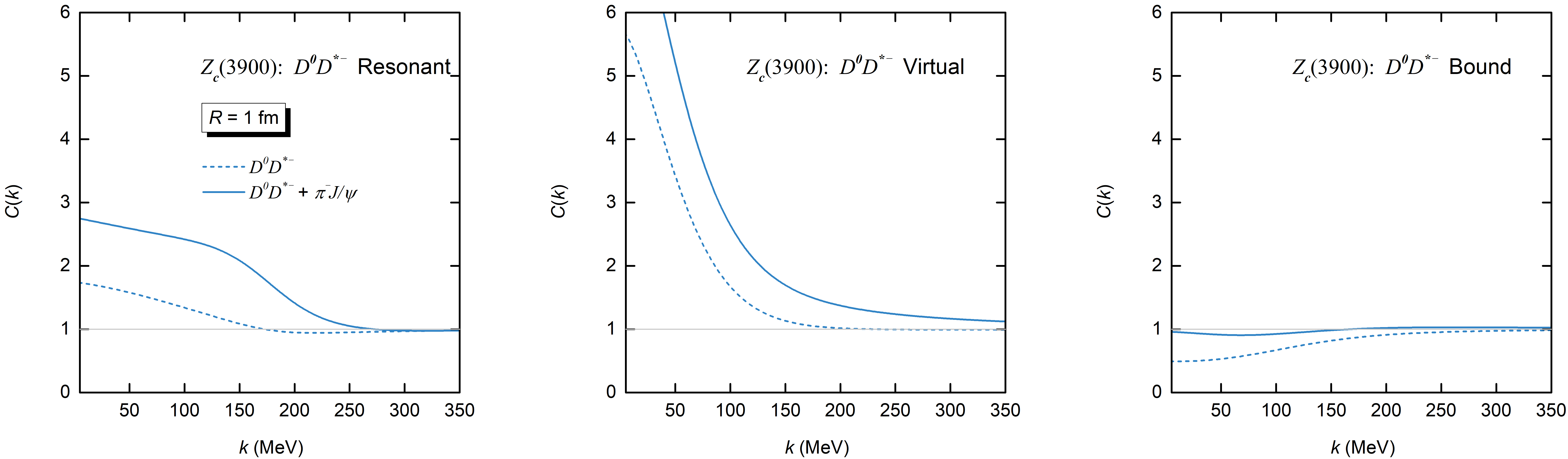}
  \caption{Coupled-channel $D^0D^{*-}$ correlation functions for $R = 1$ fm as a function of the relative momentum $k$. The dashed and solid lines denote the elastic contribution and full coupled-channel result, respectively. The sharp cutoff $q_{\rm max}$ is fixed at $1$ GeV.}\label{Fig:CF_Zc_cc}
\end{figure*}

We present the coupled-channel $D^0D^{*-}$ CFs for the three scenarios in Fig.~\ref{Fig:CF_Zc_cc}. It is observed that the elastic parts of CFs are similar to those obtained from the single-channel calculations, but full coupled-channel results are enhanced in all three scenarios. Especially in the resonant state scenario, the suppression in the high-momentum region is offset by the coupled-channel effect. This significant coupled-channel effect is mainly caused by the large $\omega_{\pi^-J/\psi}$. But considering the probability of binding $c\bar{c}$ should be less than that of binding $c\bar{q}$ or $\bar{c}q$, the realistic value of $\omega_{\pi^-J/\psi}$ may not be so large. In any case, one can still distinguish the three scenarios by measuring the significant difference in the low-momentum $D^0D^{*-}$ CFs.

\subsection{Effect of the uncertainty of pole positions}
To check the effect of pole positions on the correlation functions, we adopt the uncertainties of the latest PDG averages for the resonant state and obtain the corresponding uncertainties for the virtual and bound states demanding similar line shapes of the invariant mass spectrum, e.g., $Z_c(3900)$: resonant state $M=3887.1\pm2.6$ MeV and $\Gamma=28.4\pm2.6$ MeV~\cite{ParticleDataGroup:2024cfk}, virtual state $M=3871^{+0.5}_{-0.7}$ MeV, bound state $M=3867^{+1.8}_{-1.2}$ MeV; $Z_{cs}(3985)$: resonant state $M=3988\pm5$ MeV and $\Gamma=14\pm4$ MeV~\cite{ParticleDataGroup:2024cfk}, virtual state $M=3974^{+1.3}_{-1.0}$ MeV, bound state $M=3972^{+2.5}_{-2.2}$ MeV.

\begin{figure*}[htbp]
  \centering
  \includegraphics[width=0.96\textwidth]{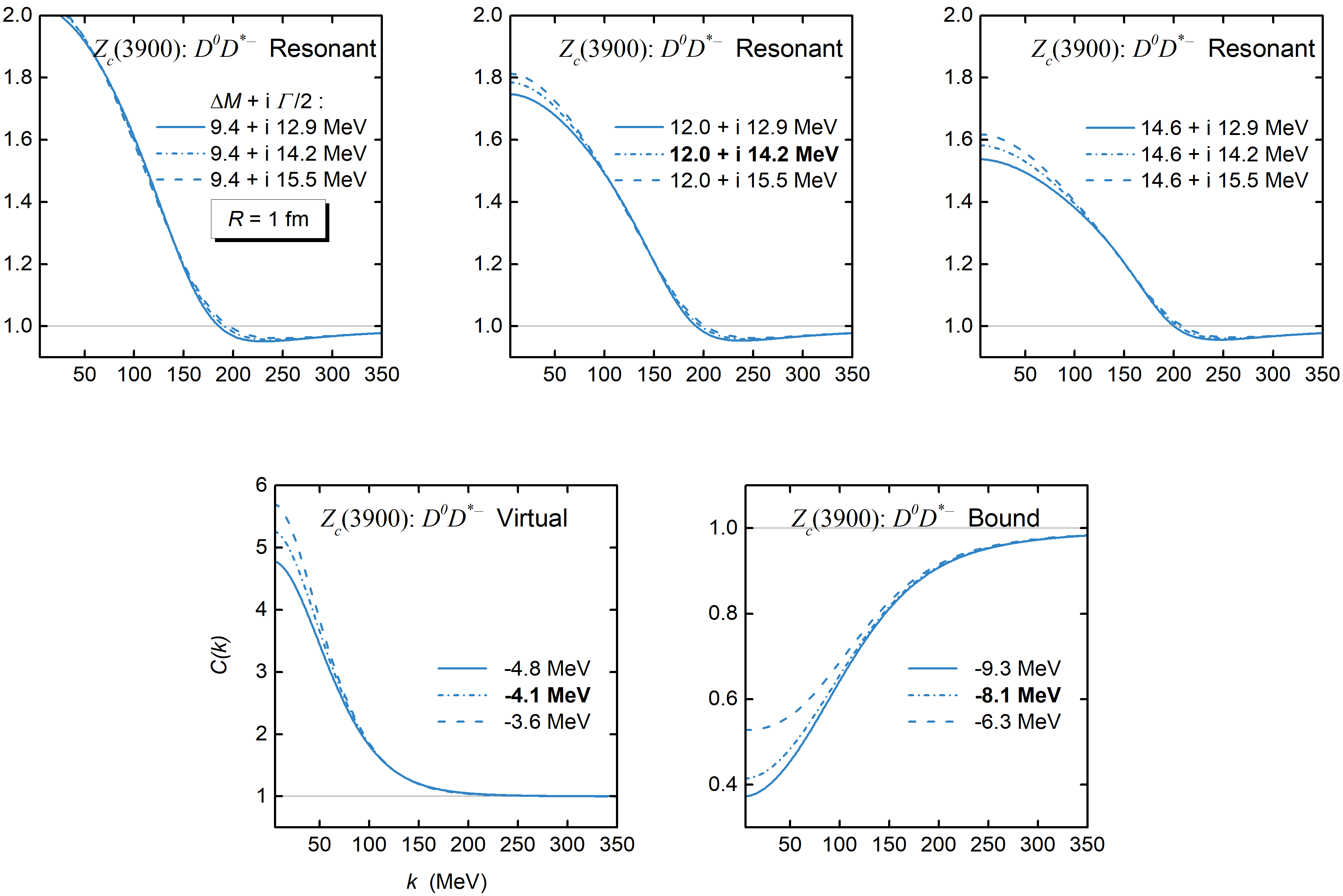}
  \caption{Pole-position dependence of the $D^0D^{*-}$ correlation functions. The source size is set at $R = 1$ fm, and the hard cutoff $q_{\rm max}$ is fixed at $1$ GeV.}\label{Fig:CF_Zc_pole}
\end{figure*}

\begin{figure*}[htbp]
  \centering
  \includegraphics[width=0.96\textwidth]{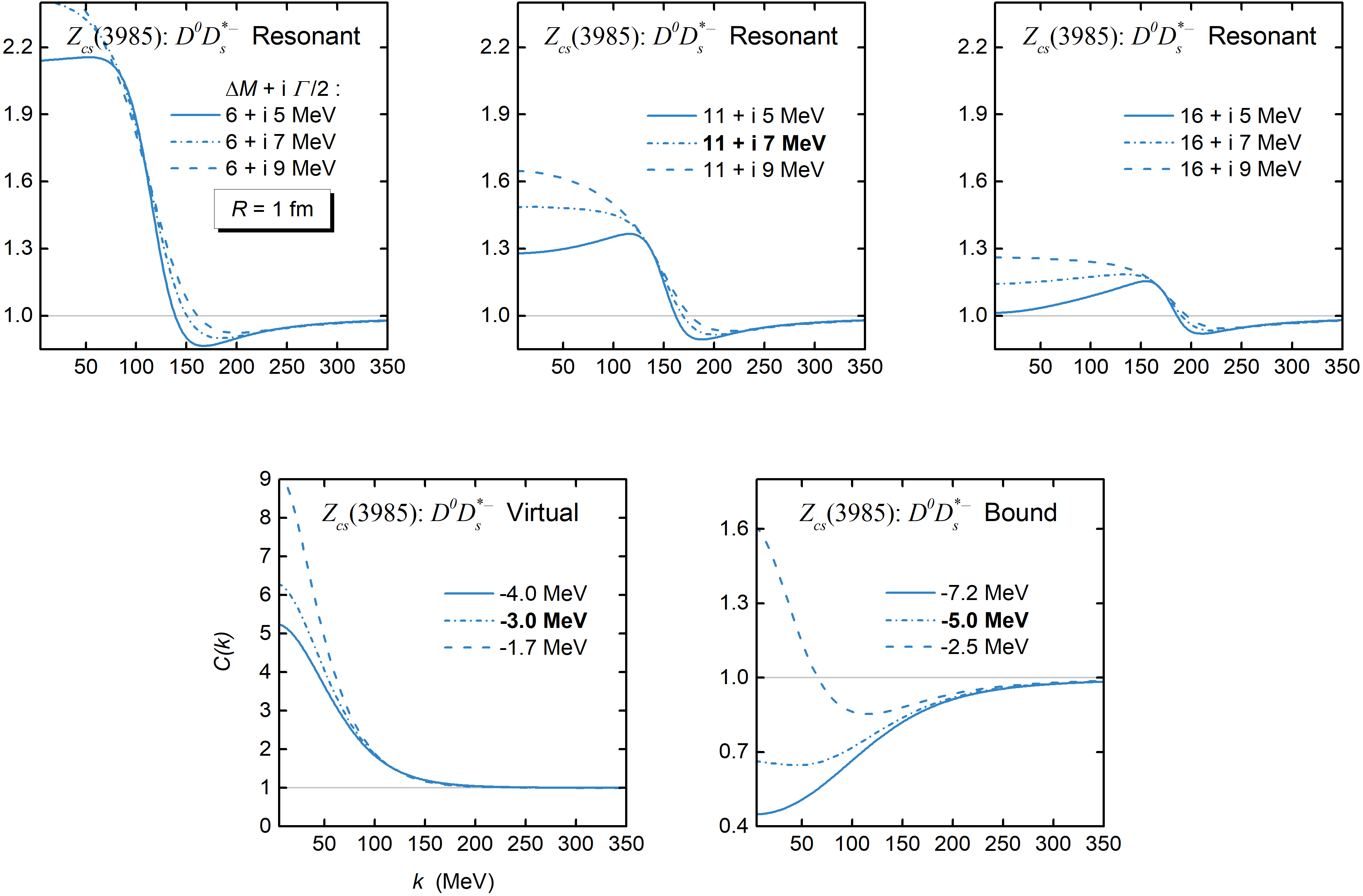}
  \caption{Pole-position dependence of the $D^0D_s^{*-}$ correlation functions. The source size is set at $R = 1$ fm, and the hard cutoff $q_{\rm max}$ is fixed at $1$ GeV.}\label{Fig:CF_Zcs_pole}
\end{figure*}

We show the pole-position dependence of the $D^0D^{*-}$ and $D^0D_s^{*-}$ CFs in Figs.~\ref{Fig:CF_Zc_pole} and \ref{Fig:CF_Zcs_pole}, respectively. For the resonant state scenario (especially for $Z_{cs}(3985)$), the peaks and minima near the resonant energy gradually evolve into low-momentum enhancement and high-momentum suppression as $\Gamma/2$ increases. This structure gradually shifts towards the high-momentum region and diminishes as $\Delta M=M-m_1-m_2$ increases. For the virtual state scenario, the CFs are always greater than unity and decrease as the virtual-state pole moves away from the threshold. For the deep bound state scenario, the CFs are always smaller than unity, and this suppression becomes more obvious with increasing binding energy. It is worth noting that in the bound state scenario of $Z_{cs}(3985)$, the CF at $\Delta M = -2.5$ MeV exhibits the feature of a shallow bound state, namely, the low-momentum result is above unity for small $R$. These general features for the bound state scenario have been demonstrated based on the square-well model in our previous work~\cite{Liu:2023uly}. In summary, it is seen that even accounting for the uncertainties of pole positions, the femtoscopic CFs can still distinguish whether $Z_c(3900)$ and $Z_{cs}(3985)$ are resonance, virtual, or bound states.

\subsection{General features of CFs and comparison with the Lednick${\rm \acute{Y}}$-Lyuboshits model}
In the following, we review the Lednick${\rm \acute{y}}$-Lyuboshits (LL) model~\cite{Lednicky:1981su}, which has been widely used in the experimental and theoretical studies of CFs~\cite{ExHIC:2017smd, Fabbietti:2020bfg}. For systems with two non-identical particles and without Coulomb interaction,
the CF can be expressed as
\begin{align}\label{Eq:KP}
C(k)=1+\int\limits_0^\infty{\rm d}^3r~S_{12}~\left[\left|\psi\right|^2-\left|j_0\right|^2\right],
\end{align}
where $\psi$ is the $S$-channel wave function. Assuming a Gaussian source $S_{12}=\exp\left[-r^2/(4R^2)\right]/(2\sqrt{\pi}R)^3$ and replacing $\psi$ everywhere with the asymptotic form 
\begin{align}\label{Eq:appWF}
  \psi(r)\approx\psi_{\rm asy}(r)=e^{-2i\delta}\left[\frac{\sin kr}{kr}+f\frac{e^{ikr}}{r}\right],
\end{align}
where $\delta$ is the phase shift and $f$ is the scattering amplitude, the CF becomes
\begin{align}\label{Eq:LL_I}
  C_{\rm LL-I}(k)=1+\frac{\left|f\right|^2}{2R^2}+\frac{2{\rm Re}f}{\sqrt{\pi}R}F_1-\frac{{\rm Im}f}{R}F_2,
\end{align}
where $F_1=\int_0^x~{\rm d}t~\exp(t^2-x^2)/x$, $F_2=[1-\exp(-x^2)]/x$, and $x=2kR$. Here, $f$ is given by the effective range expansion (ERE) $f\approx(-1/a_0+r_{\rm eff}k^2/2-ik)^{-1}$. In the low-momentum region, the deviation from the asymptotic wave function can be evaluated by using the effective range formula~\cite{Ohnishi:2016elb}
\begin{align}\label{Eq:ERF}
  \lim_{k\rightarrow0}~\frac{1}{\left|f\right|^2}\int_0^\infty r^2{\rm d}r~\left[\left|\psi\right|^2-\left|\psi_{\rm asy}\right|^2\right]=-\frac{1}{2}r_{\rm eff}.
\end{align}
When we insert a factor $\exp\left[-r^2/(4R^2)\right]$ within the integral of Eq.~\eqref{Eq:ERF}, the resulting integral can serve as the correction for Eq.~\eqref{Eq:LL_I}. This factor is expected to have an insignificant impact on the integral result as long as $R$ is large compared with the interaction range. Based on this assumption, we arrive at the CF in the so-called LL model
\begin{align}\label{Eq:LL_II}
  C_{\rm LL-II}(k)&=1+\frac{\left|f\right|^2}{2R^2}+\frac{2{\rm Re}f}{\sqrt{\pi}R}F_1-\frac{{\rm Im}f}{R}F_2\nonumber\\
  &-\frac{\left|f\right|^2}{4\sqrt{\pi}R^3}r_{\rm eff},
\end{align}
where the last term represents the effective range correction. Generally, the ERE is valid near the threshold. Away from the threshold, the ERE may break down. For instance, with $a_0 = 0.18$ fm and $r_{\rm eff} = -17.77$ fm, the scattering amplitude predicts a bound state pole below the threshold. However, this finding contradicts the resonant state pole of $\Delta M + i\Gamma/2 = 10 + i1$ MeV that was introduced initially. To avoid this problem, we adopt an alternative representation of the scattering amplitude $f=(k\cot\delta-ik)^{-1}$. Then, Eq.~\eqref{Eq:LL_I} can be rewritten in the following form
\begin{align}\label{Eq:LL_III}
  C_{\rm LL-III}(k)&=1+\left[\exp\left(-x^2\right)+\frac{2xF_1}{\sqrt{\pi}}\cot\delta\right]\frac{2}{x^2}\nonumber\\
  &\times\frac{1}{1+\cot^2\delta},
\end{align}
where we resort to the non-relativistic Breit-Wigner shape for $\cot\delta=-k_R/k\cdot(k^2-k_R^2)/(2\mu)\cdot2/\Gamma$ to capture the general features of $f$ at the resonant energy~\cite{Albaladejo:2023pzq}, where $\mu$ is the reduced mass of the hadron pair, $k_R$ is the resonant momentum corresponding to the resonance mass. In the following analyses, we refer to Eqs.~\eqref{Eq:LL_I}, \eqref{Eq:LL_II}, and \eqref{Eq:LL_III} as the LL-I, LL-II, and LL-III models, respectively.

\begin{table*}[htpb]
\centering
\setlength{\tabcolsep}{23.6pt}
\caption{Resonant, virtual, and bound state pole positions with respect to the $m_1+m_2$ threshold, the values of the parameters $a$ and $b$, and the scattering length $a_0$ and effective range $r_{\rm eff}$.\footnote{The sharp cutoff $q_{\rm max}$ is fixed at $1$ GeV.}}\label{Tab:T4}
\begin{tabular}{cccccc}
\hline
& $\Delta M+i~\Gamma/2$  & \multicolumn{1}{c}{$a$}  & \multicolumn{1}{c}{$b$}  & \multicolumn{1}{c}{$a_0$}  & \multicolumn{1}{c}{$r_{\rm eff}$}  \\
                  & (MeV)  &  & ($10^{-6}\cdot$ MeV$^{-2}$)  & \multicolumn{1}{c}{(fm)}  & \multicolumn{1}{c}{(fm)}  \\ \hline
\multirow{6}{*}{Resonant} & $2.5+i~1$  & $-108.20$  & $-9234.37$  & $-0.72$  & $-29.43$  \\
                  & $5+i~1$  & $-29.21$  & $-12934.30$  & $-0.07$  & $-570.77$  \\
                  & $10+i~1$  & $195.34$  & $-18219.00$  & $0.18$  & $-17.77$  \\
                  & $2.5+i~10$  & $-127.67$  & $-1260.96$  & $-1.43$  & $-2.62$  \\
                  & $5+i~10$  & $-121.38$  & $-1439.80$  & $-1.12$  & $-3.39$  \\
                  & $10+i~10$  & $-102.39$  & $-1796.48$  & $-0.61$  & $-6.16$  \\ \hline
\multirow{3}{*}{Virtual} & $-2.5$  & $-139.94$  & $0$  & $-2.73$  & $0.29$  \\
                  & $-5$  & $-133.68$  & $0$  & $-1.89$  & $0.29$  \\
                  & $-10$  & $-125.38$  & $0$  & $-1.30$  & $0.29$  \\ \hline
\multirow{3}{*}{Bound} & $-2.5$  & $-175.32$  & $0$  & $3.01$  & $0.28$  \\
                  & $-5$  & $-183.81$  & $0$  & $2.18$  & $0.28$  \\
                  & $-10$  & $-196.56$  & $0$  & $1.59$  & $0.28$  \\
\hline
\end{tabular}
\end{table*}

\begin{figure*}[htbp]
  \centering
  \includegraphics[width=1\textwidth]{CF_res_vir_bou_LL.png}
  \caption{Correlation functions corresponding to potentials capable of generating a resonance ($a$ and $b$ panels), virtual state ($c$ panels), or bound state ($d$ panels) with the pole position $\Delta M + i\Gamma/2$ specified in the respective plots. The source size is set at $R = 1$ and $2$ fm, and the hard cutoff $q_{\rm max}$ is fixed at $1$ GeV. The results obtained with the LL-I, LL-II, and LL-III models are shown as red dash-dotted, dashed, and dash-dot-dotted lines for comparison, respectively. For more information, please refer to the main text.}\label{Fig:CF_res_vir_bou_LL}
\end{figure*}

Without loss of generality and focusing on the charm sector, the masses of the two particles, $m_1$ and $m_2$, are chosen as 1.8 and 2 GeV, respectively, which are close to the masses of $D$ and $\bar{D}^*$. For the resonant state case, the real part of the pole position with respect to the threshold $\Delta M=M-m_1-m_2$ is set at 2.5, 5, and 10 MeV, respectively; the imaginary part of the pole position (half-width) $\Gamma/2$ is set at 1 and 10 MeV, respectively. For the virtual and bound state cases, $\Delta M$ is set at $-2.5$, $-5$, and $-10$ MeV, respectively. The parameters $a$ and $b$ are required to generate the above poles, and the corresponding scattering parameters are shown in Table~\ref{Tab:T4}.

The corresponding CFs are displayed in Fig.~\ref{Fig:CF_res_vir_bou_LL}. For a resonant state with a narrow width ((a-1)-(a-3) panels), the CF exhibits a peak at momentum slightly below $k_R$ and a valley at momentum slightly above $k_R$. However, the peak evolves into an overall enhancement in the low-momentum region for a resonant state with a larger width ((b-1)-(b-3)). In contrast, the valley in the high-momentum region becomes less pronounced. In addition, the enhancement and suppression weaken as $k_R$ moves toward the high-momentum region. For the virtual state case ((c-1)-(c-3)), the CFs are always greater than unity for various source sizes, indicating the existence of a purely attractive interaction. Meanwhile, the attractive strength decreases as the virtual-state pole moves away from the threshold, reducing the CF. This behavior is consistent with the variation of the parameter $a$ and the scattering length as a function of the pole position. For the deep bound state case ((d-2)-(d-3)), the CFs are always smaller than unity, and this suppression becomes more obvious with increasing binding energy. For the shallow bound state case (d-1), the low-momentum result is above unity for small $R$ while below unity for large $R$.

Next, we compare the exact CFs and LL model results for $R = 1$ fm, where the scattering parameters and the resonant pole positions listed in Table~\ref{Tab:T4} are employed in the LL-I/LL-II and LL-III models, respectively. For the virtual and bound state cases (see panels (c-1)-(c-3) and (d-1)-(d-3)), the exact CFs can be reproduced by the LL-II model, which is attributed to the sufficiently small effective ranges $r_{\rm eff}$ compared to the source size. However, for the resonant state case, the values of $r_{\rm eff}$ are significantly larger than the source size, which means that the effective range correction may not be appropriate. Thus, we compare the exact CFs with the LL-I results for the broad resonant state case. As shown in panels (b-1)-(b-3), the LL-I model can approximate the exact CFs. However, as the resonance moves away from the threshold, the ERE gradually breaks down, and the difference between the LL-I results and the exact CFs increases. For the narrow resonant state case (see panels (a-1)-(a-3)), since the ERE has lost the feature of a resonant state, we have to compare the exact CFs with the LL-III results. It is seen that the LL-III results may deviate from the exact results at zero momentum, but they behave similarly to the exact CFs near the resonant energy.

In addition, we also compare the exact $D^0D^{*-}$ and $D^0D_s^{*-}$ CFs with the results from the LL model. Similarly, the LL-II model can reproduce the CFs for virtual and bound states. However, for the resonant state scenarios where $|r_{\rm eff}|$ is much larger than the source size, the LL-II model breaks down due to substantial deviations introduced by the effective range correction. Meanwhile, the LL-I model performs well in reproducing CFs in the low-momentum region, whereas the LL-III model can reproduce results in the high-momentum region.

\end{document}